\documentclass[
aps,%
12pt,%
final,%
notitlepage,%
oneside,%
onecolumn,%
nobibnotes,%
nofootinbib,%
superscriptaddress,%
noshowpacs]%
{revtex4}
\usepackage{psfrag}
\usepackage{graphicx}
\usepackage{color}

\usepackage{ragged2e}
\begin{document}

% \noindent {\it Astronomy Reports, 2021, Vol. 98, No. 1}
\bigskip\bigskip  \hrule\smallskip\hrule
\vspace{35mm}

\title{The Influence of the Peculiar Velocities of SNe~Ia in Clusters of Galaxies on the Redshift Measurements}

\author{\bf
\quad \firstname{E.~A.}~\surname{Balakina,}}
\email{elena@balakin.ru}
\affiliation{Lomonosov Moscow State University, Faculty of Physics, Leninskie Gory, 1-2, Moscow, 119991, Russia}

\author{\bf \firstname{M.~V.}~\surname{Pruzhinskaya}}
\email{pruzhinskaya@gmail.com}
\affiliation{Lomonosov Moscow State University, Sternberg Astronomical Institute, Universitetsky pr.~13, Moscow, 119234, Russia}
%%%%%%%%%%%%%%%%%%%%%%%%%%%%%%%%%%%%%%%%%%%%%%%%%%%%%%%%%%%%%
\begin{abstract}
% \centerline{\footnotesize Received: ;$\;$
% Revised: ;$\;$ Accepted: .}
\bigskip\bigskip\bigskip
The most intriguing question of modern astronomy is the question of our Universe formation. The Hubble diagram analysis with Type~Ia Supernovae (SNe~Ia) is widely used to estimate the cosmological parameters with high accuracy. The cosmological measurements allow us to better understand the early stages of the Universe evolution. Nevertheless, the redshift on the SN~Ia Hubble diagram includes the contribution from the unknown peculiar velocities that decreases the accuracy of such measurements. We consider the contamination of peculiar velocities of the host galaxies for those SNe that exploded in galaxy clusters. For this purpose, we use the \textsc{Pantheon} cosmological sample and offer the procedure to minimise the influence of such uncertainties. 
\end{abstract}

\maketitle
%%%%%%%%%%%%%%%%%%%%%%%%%%%%%%%%%%%%%%%%%%%%%%%%%%%%%%%%%%%%%
\section{Introduction}
The discovery of the accelerating expansion of the Universe was performed by observations of distant  Type~Ia Supernovae (SNe~Ia)~\cite{1999Perlmutter, 1998Riess} and caused a resurgence of interest for this type of astrophysical transients. It turns out that in the last two decades the discovery rate of SNe~Ia has been dramatically increased. The future projects such as the Vera~C. Rubin Observatory Legacy Survey of Space and Time~\cite{2009arXiv0912.0201L} will provide even more observations of this phenomenon and allow us to measure the cosmological parameters with the highest accuracy.

Cosmological parameters are estimated from the ``luminosity distance-redshift'' relation of SNe~Ia using the Hubble diagram. Currently, a lot of attention is paid to enhance the accuracy of SN luminosity standardisation~\cite{2020A&A...636A..46L}. The other effect that decreases the accuracy of the luminosity distance measurements is the gravitational lensing. In~\cite{2018IJMPD..2750019H} the authors showed that its contribution at the low redshifts is negligible but increases with the redshift growth.

\begin{figure}
    \includegraphics[width=0.47\textwidth]{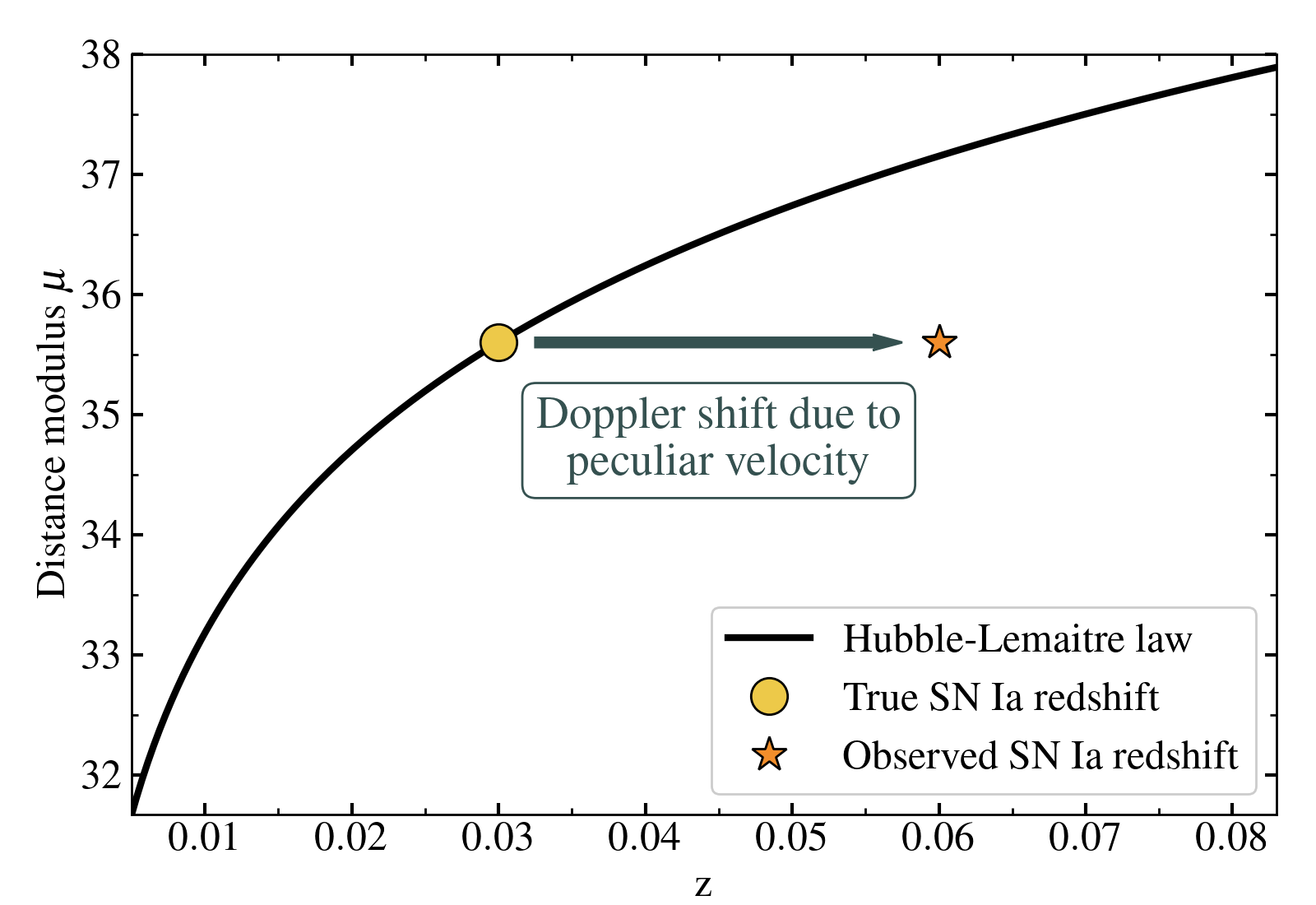}
    \includegraphics[width=0.5\textwidth]{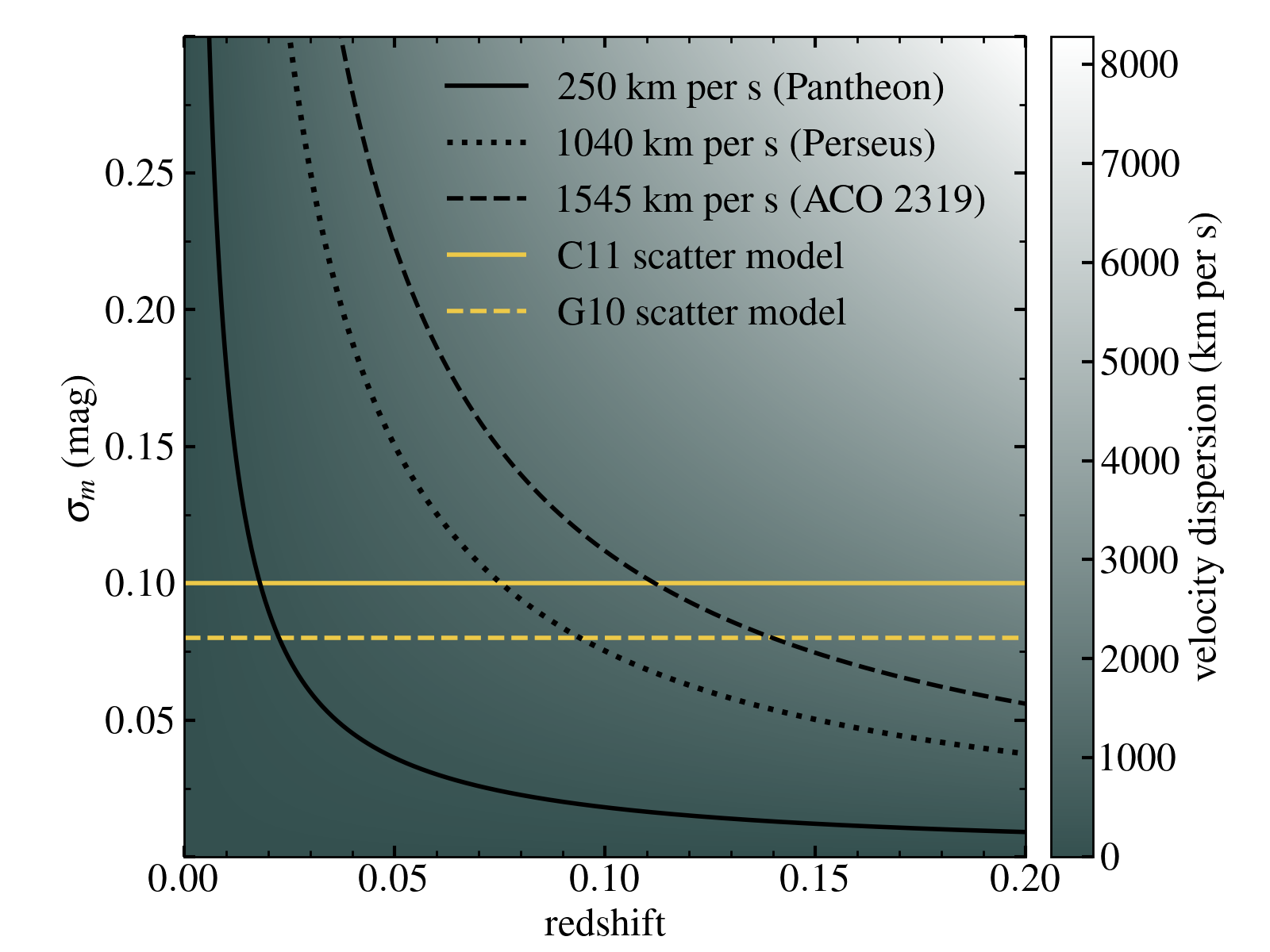}
    \caption{Left: Shift caused by the redshift uncertainty due to peculiar velocity contribution $z_{pec}$ on the Hubble diagram. Right: Redshift uncertainties due to peculiar velocities recalculated in magnitude units as a function of the cosmological redshift $z$. The solid black line corresponds to the uncertainties due to peculiar velocity correction applied in the \textsc{Pantheon} analysis; the dashed and dotted lines correspond to ACO~2319~\cite{1996ApJ...473..670F} and Perseus cluster~\cite{2020MNRAS.494.1681A} velocity dispersion, respectively. The contribution of peculiar velocities decreases with greater redshift.}
    \label{fig:shift}
\end{figure}

The uncertainty on the redshift in the ``luminosity distance-redshift'' relation is quite often considered negligible. In this relation, the redshift $z$ should be entirely due to the expansion of the Universe, i.~e. relative to the CMB frame. However, the observed redshift $z_{obs}$ contains a contribution from the poorly known peculiar velocities: $(1+z_{obs}) = (1+z)(1+z_{pec})$, where $z_{pec}$ corresponds to the peculiar velocity $v_{pec}$ contribution to the redshift. For the low redshift it could be expressed as $z_{pec} = v_{pec}/c$, where $c$ is speed of light. In Fig.~\ref{fig:shift}~(left) the impact of the peculiar velocities on the Hubble diagram is schematically shown. 

To minimise the influence of peculiar velocities, in cosmological analyses a standard value of $250$~km~s$^{-1}$ is added in quadrature to the redshift uncertainty~\cite{2018Scolnic}. It has nonetheless been observed that velocity dispersion can exceed $1000$~km~s$^{-1}$ in galaxy clusters: the velocity dispersion equals $1040$~km~s$^{-1}$ for the Perseus cluster~\cite{2020MNRAS.494.1681A} and $1545$~km~s$^{-1}$ for ACO~2319~\cite{1996ApJ...473..670F}.

One can recalculate the velocity dispersion caused by the peculiar velocities to the magnitude uncertainty by $\sigma_m = \frac{5 \sigma_V}{c z \ln{10}}$. For instance, if a SN~Ia host galaxy at redshift $z = 0.05$ has a peculiar velocity of $1040$~km~s$^{-1}$, its contribution to the magnitude uncertainty will be $\sigma_m \simeq 0.15$~mag. As it is seen from Fig.~\ref{fig:shift}~(right), the magnitude uncertainty due to peculiar velocities of $1000-1500$~km~s$^{-1}$ is greater than the intrinsic dispersion of $\sigma_m\simeq0.11$~mag for the scattering model C11~\cite{2011A&A...529L...4C} in the \textsc{Pantheon} analysis for the redshifts $z < 0.15$ and, therefore, has to be taken into account~\cite{2020arXiv201005765H}. To decrease the contribution of this effect for those SNe that exploded in the galaxy clusters, we suggest to use the redshift of the clusters instead of the redshift of their host galaxies. 

The similar analysis was performed for 11 Nearby Supernova Factory SNe~Ia in galaxy clusters~\cite{2018A&A...615A.162L}. The authors showed that the Hubble residual dispersion of SNe~Ia associated with galaxy clusters significantly decreased when the cluster redshift is taken as the cosmological redshift of supernovae. Our analysis applies the similar procedure to the \textsc{Pantheon} cosmological sample of SNe~Ia~\cite{2018Scolnic}.

The paper is organised as follows. In Section~\ref{sec:data} we present the \textsc{Pantheon} data and used galaxy cluster catalogues. Section~\ref{sec:search} describes the cross-matching procedure and the criteria of supernova search in the galaxy clusters. Finally, we conclude the paper in Section~\ref{sec:conclusion}.

%%%%%%%%%%%%%%%%%%%%%%%%%%%%%%%%%%%%%%%%%%%%%%%%%%%%%%%%%%%%%
\section{Data}
\label{sec:data}
\subsection{\textsc{Pantheon} sample}
\label{subsec:pantheon}
As a supernova sample we use \textsc{Pantheon} cosmological sample consisting of 1048 spectroscopically confirmed SNe~Ia with redshift range $0.01 < z < 2.3$. This sample represents a compilation from several supernova surveys~\cite{2018Scolnic}. We choose this sample because it is one of the biggest intercalibrated samples of supernovae. Since the contribution of peculiar velocities is significant for the nearby Universe (see Fig.~\ref{fig:shift}, right) we chose 297 SNe~Ia from Pantheon sample with the redshifts $z < 0.15$.

\subsection{Galaxy clusters data}
\label{subsec:clusters}
Galaxy clusters are characterised by the extremely hot gas in their centres that is observed as the diffuse X-ray emission~\cite{1966PhRvL..17..447B, 1988xrec.book.....S}. This observational fact can serve as the most effective and reliable way to establish which optically detected galaxy clusters are  gravitationally-bound systems. Another observational appearance of hot gas in the central parts of clusters is the Sunyaev-Zel’dovich~(SZ) effect that manifests as the distortion in the cosmic microwave background. Based on these effects we collect the following catalogues with galaxy clusters:
\begin{itemize}
    \item MCXC which is the compilation of X-ray detected clusters of galaxies. This catalogue is based on publicly available ROSAT All~Sky Survey-based, serendipitous cluster catalogues and finally comprises 1743 clusters~\cite{2011MCXC}.
    \item SWXCS which is the Swift X-ray Cluster Survey catalogue obtained using archival data from the X-ray telescope onboard the Swift satellite. The list of sources was cross-correlated with published optical, X-ray, and Sunyaev-Zel’dovich catalogues of clusters and 263 sources have been identified as clusters~\cite{2015ApJS..216...28L}.
    \item XCS-DR1 which is the first data release from the XMM Cluster Survey. It consists of 503 detected X-ray clusters that were optically confirmed~\cite{2012XCS}.
    \item Plank-SZ catalogue that consists of 439 clusters detected via their Sunyaev-Zel’dovich signal~\cite{2016Planck}.
\end{itemize}

%%%%%%%%%%%%%%%%%%%%%%%%%%%%%%%%%%%%%%%%%%%%%%%%%%%%%%%%%%%%%
\section{Search of SNe~Ia in galaxy clusters}
\label{sec:search}
\subsection{Preliminary search}
\begin{figure}
\centering
	\includegraphics[width=0.60\textwidth]{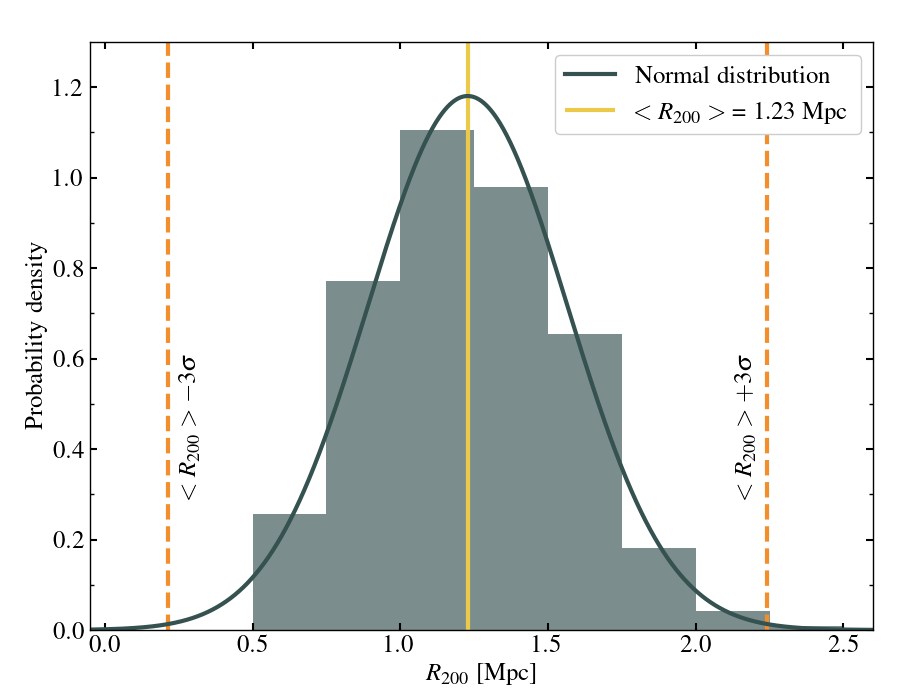}
    \caption{Virial radius $R_{200}$ distribution of galaxy clusters from MCXC~\citep{2011MCXC}. The dark green solid line corresponds to the Gauss distribution of the clusters, the yellow solid line shows the mean $R_{200}$ value, the yellow dashed lines present $3\sigma$ area for this distribution.}
    \label{fig:hist}
\end{figure}
To find those \textsc{Pantheon} SNe that belong to clusters of galaxies, we realise the cross-matching between the chosen cluster catalogues and \textsc{Pantheon} sample. The preliminary search is performed in the cone of characteristic value $R_{max}$. To find $R_{max}$, we build the distribution of $R_{200}$ parameter based on MCXC data, where $R_{200}$ is the radius that encloses a sphere whose mean density is equal to 200 times of the critical density of the Universe~(Fig.~\ref{fig:hist}). The $R_{200}$ parameter is calculated from $R_{500}$ parameter from MCXC using the relation $R_{200} \simeq 1.5 R_{500}$. After finding the mean value of $R_{200}$ and its standard deviation $\sigma_{R_{200}}$ we choose $R_{max}$ criterion as $R_{max} = <R_{200}> + 5 \sigma_{R_{200}} \simeq 2.92$~Mpc. Besides, for this preliminary search, we need to define some redshift cut. For the ACO~2319 cluster which demonstrates one of the highest value of velocity dispersion we calculate the maximum value of $\Delta z = |z_{SN} - z_{cl}| \leq 5 \sigma_{v_{pec}} / c \simeq 0.025$. This value is used for the preliminary redshift cut. After applying both cuts, 30 SNe~Ia in the galaxy clusters are identified by cross-matching procedure.

\subsection{Redshift of galaxy clusters}
\begin{figure}[h]
\centering
	\includegraphics[width=0.65\textwidth]{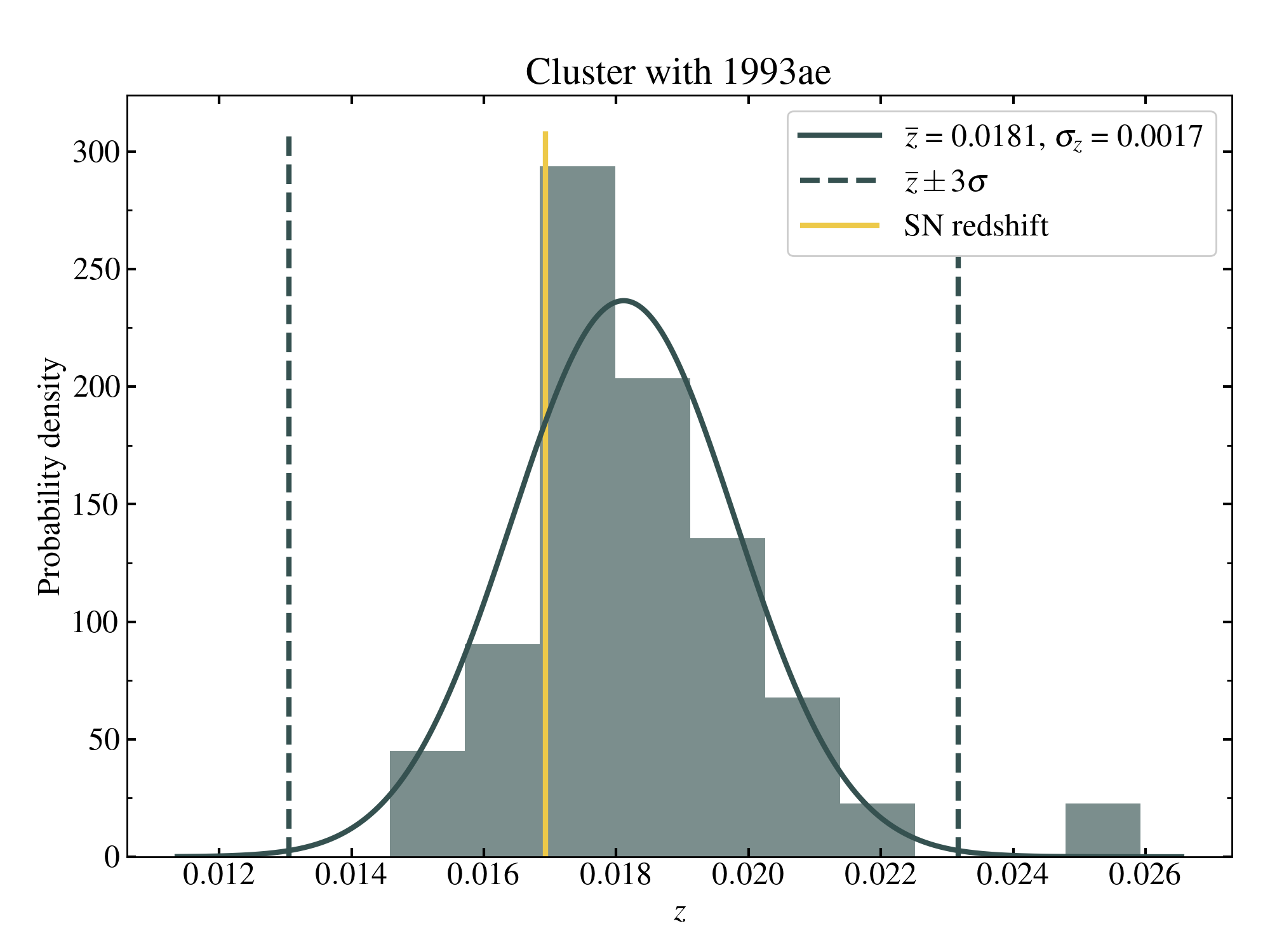}
    \caption{The redshift distribution for galaxies inside the virial radius $R_{200}$ of J0125.6-0124 cluster that hosts SN~1993ae. The supernova redshift given by~\textsc{Pantheon} is plotted by the yellow solid line. The dark green solid line corresponds to the Gauss distribution with the mean and the standard deviation values provided by the bi-weight technique. The dark green dash lines show $3\sigma_z$ area.}
    \label{fig:cluster}
\end{figure}

To reduce the residuals on the Hubble diagram, we suggest in our analysis to use the galaxy cluster redshifts instead of the host galaxy redshifts for SNe~Ia that belong to the clusters. To measure the cluster redshift, for each cluster we select all galaxies inside $R_{200}$ radius around cluster centre that have the spectroscopic redshifts in SDSS\footnote{\url{https://www.sdss.org}}~\cite{2019PASP..131e5001W,2020ApJS..249....3A}. Following~\cite{2018A&A...615A.162L} we choose the objects with the redshift value in the range $z_{cl} \pm 5\sigma_V$, where the cluster velocity dispersion $\sigma_V$ could be found as $\sigma_V \approx 10 R_{200} H(z)$, $H(z) = H_0 \sqrt{(1-\Omega_{\Lambda})(1+z)^3 + \Omega_{\Lambda}}$ for the flat Universe. For the resulting list of objects, we apply the bi-weight technique~\cite{1990AJ....100...32B} and estimate the redshift values for 25 clusters. For example, the histogram with redshift distribution for one of the clusters is shown in Fig.~\ref{fig:cluster}. For other five clusters there are no data in SDSS, and we collect their redshifts from the SIMBAD database~\cite{2000A&AS..143....9W} and literature. Thus, we define the redshifts for all 30 galaxy clusters as well as their virial radii. 

\subsection{Final selection}
After determining the redshift of galaxy clusters, we can perform the final selection to discard those supernovae that only projected near the clusters. Therefore, each SN has to satisfy the following conditions: I. $|z_{SN} - z_{cl}| < 3 \sigma_v / c$ and II. $r < R_{200}$, where $r$ is a distance between supernova and galaxy cluster centre. These final cuts allow us to determine those \textsc{Pantheon} SNe~Ia that belong to the galaxy clusters. Table~\ref{tab:final} presents the final list of 13 supernovae associated with clusters of galaxies and the main physical parameters of the clusters.

%%%%%%%%%%%%%%%%%%%%%%%%%%%%%%%%%%%%%%%%%%%%%%%%%%%%%%%%%%%%%
\begin{figure}
\centering
	\includegraphics[width=0.80\textwidth]{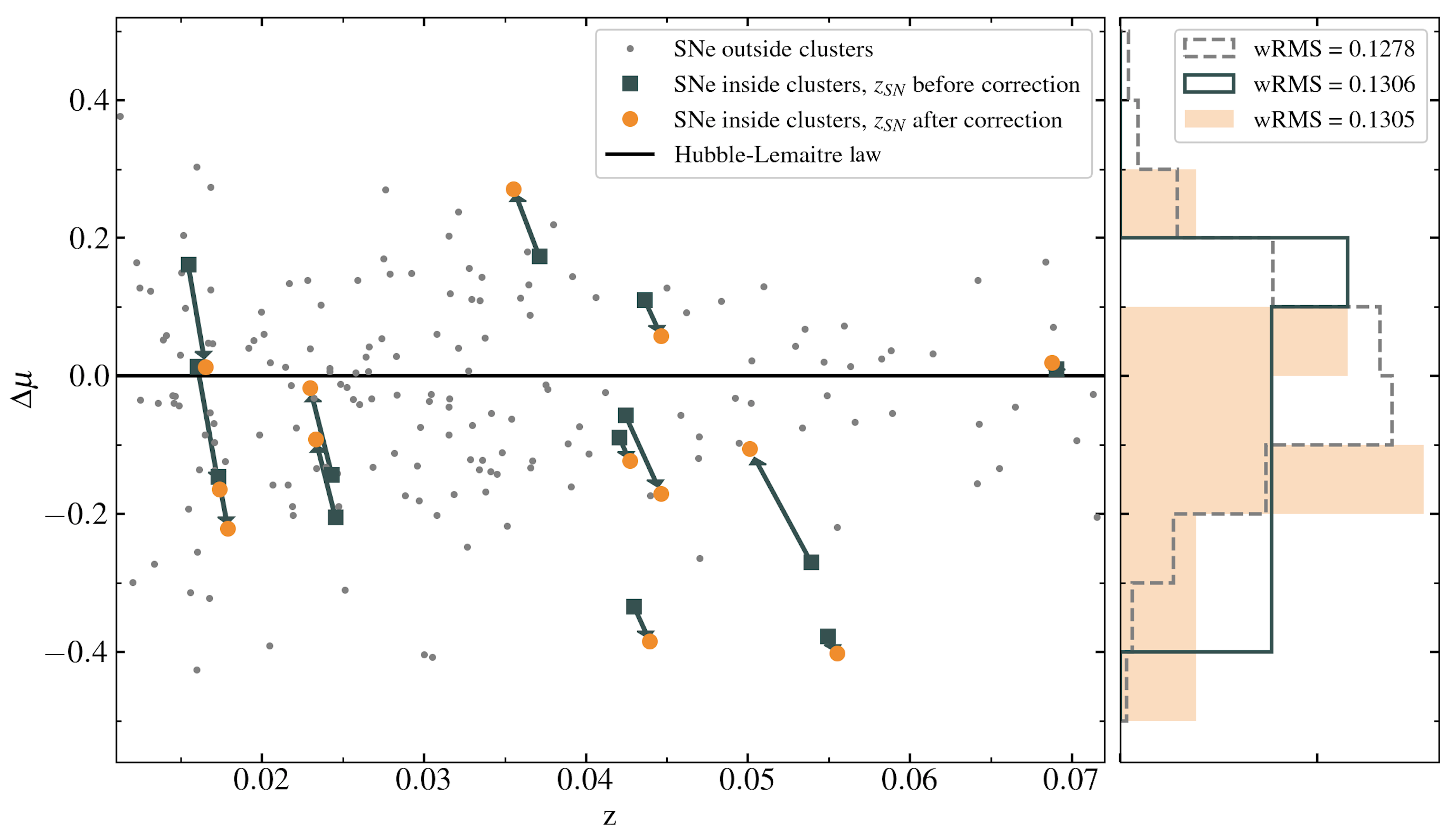}
    \caption{Distance modulus residual on the Hubble diagram before peculiar velocity correction (dark green squares) and after it (orange circles). Hubble-Lemaitre law is plotted by the black solid line based on the parameter values from~\cite{2018Scolnic}.}
    \label{fig:residuals}
\end{figure}

%%%%%%%%%%%%%%%%%%%%%%%%%%%%%%%%%%%%%%%%%%%%%%%%%%%%%%%%%%%%%
\section{Discussion}
\label{sec:discussion}
\begin{figure}
\centering
	\includegraphics[width=0.73\textwidth]{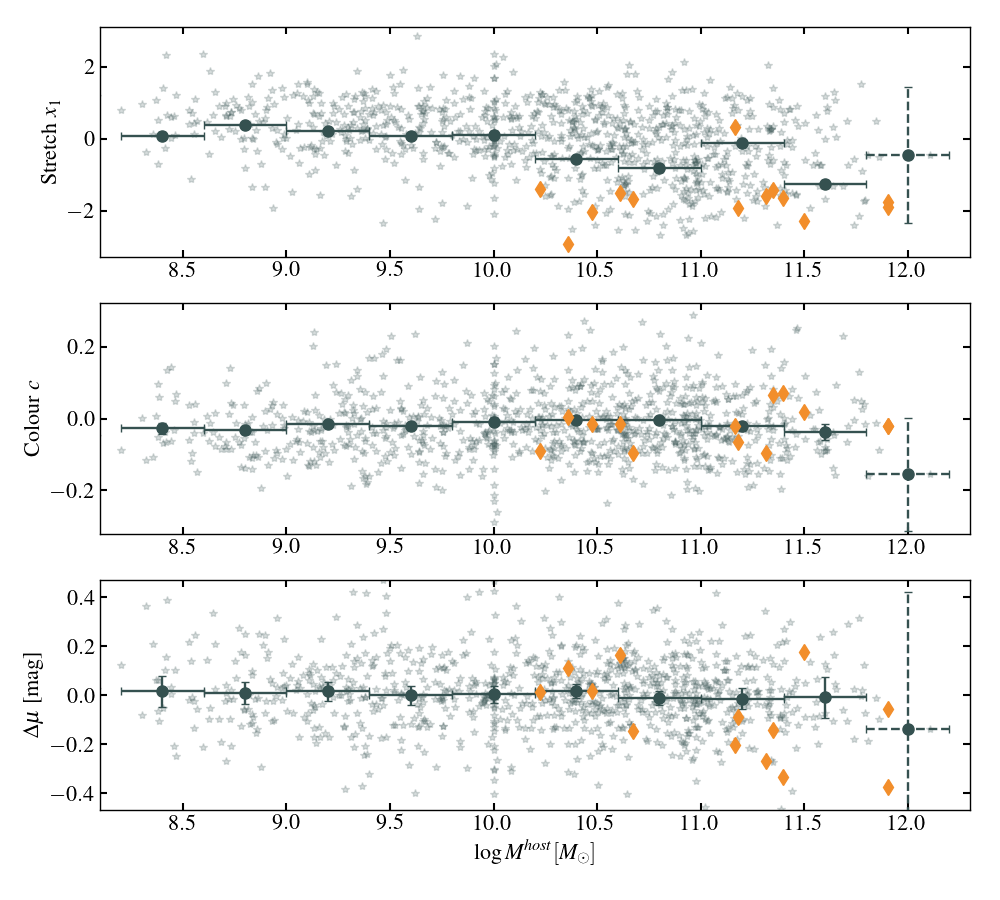}
    \caption{The dependence of the parameters $x_1$, $c$ and $\Delta \mu$ on the mass of the host galaxy for SN~Ia from~\cite{2018Scolnic}. The bars illustrate the average values of the parameters for the selected mass range of host galaxies. SNe in clusters are shown with yellow rhombuses.}
    \label{fig:mstellar}
\end{figure}

The effects of environment is another cause of uncertainties on the Hubble diagram. Using Pantheon data we consider the influence of environmental effects: the position of the SN relative to its host galaxy and the mass of the host galaxy -- for supernovae in the field and supernovae in clusters, see Fig.~\ref{fig:mstellar}. One can notices that host galaxies related with clusters are massive: $\log_{10}{M^{host}[M_{\odot}]} > 10$. Besides, SNe in clusters show even lower values of the stretching parameter $x_1$ in comparison with the average values that shown by green bars in Fig.~\ref{fig:mstellar} for each $\log_{10}{M^{host}[M_{\odot}]}$ ranges. This effect may be due to the fact that in a cluster of galaxies it is more likely to detect elliptical host galaxies, which are characterized by large masses and lower values of the stretching parameter~\cite{2020MNRAS.499.5121P}.

\section{Conclusion}
\label{sec:conclusion}
Using the \textsc{Pantheon} distance modulus values $\mu_{obs}$ and the theoretical distance modulus for the flat $\Lambda$CDM cosmology~\cite{2018Scolnic}, we found the distance modulus residuals for 13 SNe~Ia in clusters. Then, we corrected the \textsc{Pantheon} redshifts by the peculiar velocity correction found in Section~\ref{sec:search}, and compared the Hubble diagram residuals before and after corrections (see Fig.~\ref{fig:residuals}).
As expected, the supernova positions on the Hubble diagram do change. In perspective, we are going to evaluate the significance of the impact of the peculiar velocity correction on the Hubble diagram. However, it can be already stressed that with increasing accuracy of the cosmological studies, such effects become more and more important and have to be taken into account.

%%%%%%%%%%%%%%%%%%%%%%%%%%%%%%%%%%%%%%%%%%%%%%%%%%%%%%%%%%%%%
\begin{acknowledgments}
The authors are grateful to Pierre-François L{\'e}get for the helpful discussion. This work has made use of data from the Sloan Digital Sky Surveys~\cite{2020ApJS..249....3A} and the SIMBAD database, operated at CDS, Strasbourg, France~\cite{2000A&AS..143....9W}.
\end{acknowledgments}
\section*{Funding}
E.A.B. and M.V.P. acknowledge support from RSF grant 18-72-00159 and the Interdisciplinary Scientific and Educational School of Moscow University ``Fundamental and Applied Space Research". 

% \clearpage
\centerline{REFERENCES}
\bibliographystyle{ieeetr57}
\bibliography{biblio}

\clearpage
\begin{table}
\caption{The final sample of \textsc{Pantheon} SNe~Ia in galaxy clusters. The 1st column contains SN names, the 2nd one --- SN redshifts from the \textsc{Pantheon} sample. The 3rd, 4th, 5th, and 6th columns contain the names of associated galaxy clusters, their coordinates from MCXC and redshift. Parameters $R_{200}$ for each cluster are presented in the 7th column. The last column contains the references for the redshift $z_{cl}$ source.}
\label{tab:final}
\begin{tabular}{l|c|c|c|c|c|c|r}
SN & $z_{SN}$ & Cluster name & R.A.$_{cl}$ & Dec.$_{cl}$ & $z_{cl}$ & $R_{200}$ [Mpc] & Ref. \\
\hline
% 2001en~~&~~0.01544~~&~~J0123.2+3327~~&~~01 23 12.2~~&~~+33 27 40~~&~~0.0165~~&~~0.761~~&~~0.013~~&~\cite{2019PASP..131e5001W} \\
2001ic~~&~~0.04296~~&~~J2310.4+0734~~&~~23 10 26.4~~& +07 34 37~~&~~0.0439~~&~~1.117~~&~~\cite{2019PASP..131e5001W} \\
% 2002bz  & 0.03762  & J1423.1+2615 & 14 23 10.1  & +26 15 20   & 0.0376 & 0.765  & 0.009 & \cite{2019PASP..131e5001W} \\
% 2002de  & 0.02827  & J1617.4+3456 & 16 17 27.2  & +34 56 12   & 0.0309 & 0.931  & 0.014 & \cite{2019PASP..131e5001W} \\
% 2002eu  & 0.03671  & J0150.7+3305 & 01 50 44.9  & +33 05 24   & 0.0355 & 0.907  & 0.009 & \cite{2019PASP..131e5001W} \\
2003ic  & 0.05491  & J0041.8$-$0918 & 00 41 50.1  & -09 18 07   & 0.0555 & 1.840  & \cite{2019PASP..131e5001W} \\
2006al  & 0.06905  & J1039.4+0510 & 10 39 27.9  & +05 10 46   & 0.0688 & 1.225  & \cite{2019PASP..131e5001W} \\
2006gt  & 0.04362  & J0056.3$-$0112 & 00 56 18.3  & -01 13 00   & 0.0446 & 1.431  & \cite{2019PASP..131e5001W} \\
2006je  & 0.03712  & J0150.7+3305 & 01 50 44.9  & +33 05 24   & 0.0355 & 0.907 & \cite{2019PASP..131e5001W} \\
2008bf  & 0.02453  & J1204.1+2020 & 12 04 11.6  & +20 20 53   & 0.0233 & 0.727 & \cite{2019PASP..131e5001W} \\
% 2007hu  & 0.0354   & J1658.0+2751 & 16 58 00.8  & +27 51 16   & 0.0352 & 0.946  & 0.006 & \cite{2019PASP..131e5001W} \\
% 2007su  & 0.02662  & J2214.8+1350 & 22 14 52.7  & +13 50 48   & 0.0260 & 0.724  & 0.022 & \cite{2019PASP..131e5001W} \\
% 2008ar  & 0.02739  & J1223.1+1037 & 12 23 06.5  & +10 37 26   & 0.0256 & 0.877  & 0.008 & \cite{2019PASP..131e5001W} \\
% 2010A   & 0.01985  & J0231.9+0114 & 02 31 57.1  & +01 14 40   & 0.0222 & 0.657  & 0.011 & \cite{2019PASP..131e5001W} \\
2007nq  & 0.04243  & J0056.3$-$0112 & 00 56 18.3  & -01 13 00   & 0.0446 & 1.431 & \cite{2019PASP..131e5001W} \\
% 1993ae  & 0.01693  & J0125.6$-$0124 & 01 25 40.8  & -01 24 26   & 0.0181 & 0.784  & 0.021 & \cite{2019PASP..131e5001W} \\
1994M   & 0.02431  & J1231.0+0037 & 12 31 05.6  & +00 37 44   & 0.0230 & 0.621 & \cite{2019PASP..131e5001W} \\
% 1997dg  & 0.0328   & J2338.4+2700 & 23 38 25.7  & +27 00 45   & 0.0313 & 1.134  & 0.016 & \cite{2019PASP..131e5001W} \\
1999ej  & 0.01544  & J0123.2+3327 & 01 23 12.2  & +33 27 40   & 0.0165 & 0.761 & \cite{2019PASP..131e5001W} \\
2000dk  & 0.01602  & J0107.4+3227 & 01 07 28.3  & +32 27 43   & 0.0174 & 0.784 & \cite{2019PASP..131e5001W} \\
2005hf  & 0.04205  & J0126.9+1912 & 01 26 54.5  & +19 12 42   & 0.0427 & 0.519 & \cite{1999ApJS..121..287H} \\
% 2006td  & 0.01504  & J0152.7+3609 & 01 52 46.8  & +36 09 05   & 0.016  & 1.130  & 0.020 & \cite{2016MNRAS.458..264H} \\
% 2007ae  & 0.06416  & J1703.8+7838 & 17 03 48.8  & +78 38 40   & 0.0583 & 1.706  & 0.007 & \cite{2016ApJ...819...63R} \\
2008L   & 0.0173   & J0319.7+4130 & 03 19 47.2  & +41 30 47   & 0.0179 & 1.954  & \cite{2016Natur.535..117H} \\
1998dx  & 0.05389  & J1811.0+4954 & 18 11 00.1  & +49 54 40   & 0.0501 & 1.166 & \cite{2000ApJS..129..435B} \\
\end{tabular}
\end{table}

\end{document}